\documentclass[conference]{IEEEtran}
\usepackage{amsmath}
\usepackage{graphicx}
\usepackage{cite}
\usepackage{epstopdf}
\usepackage{balance}
\usepackage{gensymb}
\usepackage{color}
\usepackage{lipsum}
\usepackage{float}
\usepackage{epstopdf}
\usepackage[mathcal]{eucal}
\usepackage{subfiles}
\usepackage[T1]{fontenc}
\usepackage[unicode=true,
 bookmarks=true,bookmarksnumbered=true,bookmarksopen=true,bookmarksopenlevel=1,
 breaklinks=false,pdfborder={0 0 0},backref=false,colorlinks=false]
 {hyperref}
\hypersetup{draft}

\newcommand{\ignore}[1]{}

\usepackage{breqn}
\usepackage{booktabs}
\usepackage{multirow}

\usepackage[acronyms,nonumberlist,nopostdot,nomain,nogroupskip]{glossaries}
\newacronym{quic}{QUIC}{Quick UDP Internet Connections}
\newacronym{3gpp}{3GPP}{3rd Generation Partnership Project}
\newacronym{adc}{ADC}{Analog to Digital Converter}
\newacronym{5g}{5G}{5th generation}
\newacronym{aimd}{AIMD}{Additive Increase Multiplicative Decrease}
\newacronym{am}{AM}{Acknowledged Mode}
\newacronym{amc}{AMC}{Adaptive Modulation and Coding}
\newacronym{aqm}{AQM}{Active Queue Management}
\newacronym{awgn}{AGWN}{Additive White Gaussian Noise}
\newacronym{afd}{AFD}{Austin Fire Department}
\newacronym{balia}{BALIA}{Balanced Link Adaptation}
\newacronym{bdp}{BDP}{Bandwidth-Delay Product}
\newacronym{bf}{BF}{Beamforming}
\newacronym{cc}{CC}{Congestion Control}
\newacronym{cdf}{CDF}{Cumulative Distribution Function}
\newacronym{cn}{CN}{Core Network}
\newacronym{cqi}{CQI}{Channel Quality Information}
\newacronym{cp}{CP}{Control Plane}
\newacronym{csirs}{CSI-RS}{Channel State Information - Reference Signal}
\newacronym{dc}{DC}{Dual Connectivity}
\newacronym{dce}{DCE}{Direct Code Execution}
\newacronym{dci}{DCI}{Downlink Control Information}
\newacronym{dl}{DL}{Downlink}
\newacronym{dmr}{DMR}{Deadline Miss Ratio}
\newacronym{dmrs}{DMRS}{DeModulation Reference Signal}
\newacronym{e2e}{E2E}{End-to-End}
\newacronym{ecn}{ECN}{Explicit Congestion Notification}
\newacronym{edf}{EDF}{Earliest Deadline First}
\newacronym{enb}{eNB}{evolved Node Base}
\newacronym{epc}{EPC}{Evolved Packet Core}
\newacronym{es}{ES}{Edge Server}
\newacronym{fdma}{FDMA}{Frequency Division Multiple Access}
\newacronym{fdd}{FDD}{Frequency Division Duplexing}
\newacronym[firstplural=Radio Access Technologies (RATs)]{rat}{RAT}{Radio Access Technology}
\newacronym{fs}{FS}{Fast Switching}
\newacronym{ftp}{FTP}{File Transfer Protocol}
\newacronym{gnb}{gNB}{Next Generation Node Base}
\newacronym{harq}{HARQ}{Hybrid Automatic Repeat reQuest}
\newacronym{hetnet}{HetNet}{Heterogeneous Network}
\newacronym{hh}{HH}{Hard Handover}
\newacronym{hol}{HOL}{Head-of-Line}
\newacronym{ia}{IA}{Initial Access}
\newacronym{imt}{IMT}{International Mobile Telecommunication}
\newacronym{iot}{IoT}{Internet of Things}
\newacronym{los}{LOS}{Line of Sight}
\newacronym{lte}{LTE}{Long Term Evolution}
\newacronym{m2m}{M2M}{Machine to Machine}
\newacronym{mac}{MAC}{Medium Access Control}
\newacronym{mc}{MC}{Multi-Connectivity}
\newacronym{mcs}{MCS}{Modulation and Coding Scheme}
\newacronym{mec}{MEC}{Mobile Edge Cloud}
\newacronym{mi}{MI}{Mutual Information}
\newacronym{mimo}{MIMO}{Multiple Input, Multiple Output}
\newacronym{mmwave}{mmWave}{millimeter wave}
\newacronym{mr}{MR}{Maximum Rate}
\newacronym{mss}{MSS}{Maximum Segment Size}
\newacronym{mtd}{MTD}{Machine-Type Device}
\newacronym{mtu}{MTU}{Maximum Transmission Unit}
\newacronym{nfv}{NFV}{Network Function Virtualization}
\newacronym{nlos}{NLOS}{Non Line of Sight}
\newacronym{nr}{NR}{New Radio}
\newacronym{ofdm}{OFDM}{Orthogonal Frequency Division Multiplexing}
\newacronym{pdcch}{PDCCH}{Physical Downlonk Control Channel}
\newacronym{pdcp}{PDCP}{Packet Data Convergence Protocol}
\newacronym{pdsch}{PDSCH}{Physical Downlink Shared Channel}
\newacronym{pdu}{PDU}{Packet Data Unit}
\newacronym{pf}{PF}{Proportional Fair}
\newacronym{pgw}{PGW}{Packet Gateway}
\newacronym{phy}{PHY}{Physical}
\newacronym{pbch}{PBCH}{Physical Broadcast Channel}
\newacronym[plural=\gls{mme}s,firstplural=Mobility Management Entities (MMEs)]{mme}{MME}{Mobility Management Entity}
\newacronym{prb}{PRB}{Physical Resource Block}
\newacronym{pss}{PSS}{Primary Synchronization Signal}
\newacronym{pucch}{PUCCH}{Physical Uplink Control Channel}
\newacronym{pusch}{PUSCH}{Physical Uplink Shared Channel}
\newacronym{rach}{RACH}{Random Access Channel}
\newacronym{ran}{RAN}{Radio Access Network}
\newacronym{red}{RED}{Robotics Emergency Deployment}
\newacronym{rf}{RF}{Radio Frequency}
\newacronym{rlc}{RLC}{Radio Link Control}
\newacronym{rlf}{RLF}{Radio Link Failure}
\newacronym{rrc}{RRC}{Radio Resource Control}
\newacronym{rrm}{RRM}{Radio Resource Management}
\newacronym{rr}{RR}{Round Robin}
\newacronym{rs}{RS}{Remote Server}
\newacronym{rsrp}{RSRP}{Reference Signal Received Power}
\newacronym{rss}{RSS}{Received Signal Strength}
\newacronym{rtt}{RTT}{Round Trip Time}
\newacronym{rw}{RW}{Receive Window}
\newacronym{rx}{RX}{Receiver}
\newacronym{sa}{SA}{standalone}
\newacronym{sack}{SACK}{Selective Acknowledgment}
\newacronym{sap}{SAP}{Service Access Point}
\newacronym{sch}{SCH}{Secondary Cell Handover}
\newacronym{scoot}{SCOOT}{Split Cycle Offset Optimization Technique}
\newacronym{sdma}{SDMA}{Spatial Division Multiple Access}
\newacronym{sinr}{SINR}{Signal to Interference plus Noise Ratio}
\newacronym{sm}{SM}{Saturation Mode}
\newacronym{snr}{SNR}{Signal to Noise Ratio}
\newacronym{son}{SON}{Self-Organizing Network}
\newacronym{ss}{SS}{Synchronization Signal}
\newacronym{srs}{SRS}{Sounding Reference Signal}
\newacronym{sss}{SSS}{Secondary Synchronization Signal}
\newacronym{tb}{TB}{Transport Block}
\newacronym{tcp}{TCP}{Transmission Control Protocol}
\newacronym{tdd}{TDD}{Time Division Duplexing}
\newacronym{tdma}{TDMA}{Time Division Multiple Access}
\newacronym{tfl}{TfL}{Transport for London}
\newacronym{tm}{TM}{Transparent Mode}
\newacronym{trp}{TRP}{Transmitter Receiver Pair}
\newacronym{tti}{TTI}{Transmission Time Interval}
\newacronym{ttt}{TTT}{Time-to-Trigger}
\newacronym{tx}{TX}{Transmitter}
\newacronym{ue}{UE}{User Equipment}
\newacronym{ul}{UL}{Uplink}
\newacronym{uml}{UML}{Unified Modeling Language}
\newacronym{um}{UM}{Unacknowledged Mode}
\newacronym{utc}{UTC}{Urban Traffic Control}
\newacronym{vm}{VM}{Virtual Machine}
\newacronym{rsrq}{RSRQ}{Reference Signal Received Quality}
\newacronym{rssi}{RSSI}{Received Signal Strength Indicator}
\newacronym{crs}{CRS}{Cell Reference Signal}
\newacronym{comp}{CoMP}{Coordinated Multi-Point}
\newacronym{cran}{C-RAN}{Cloud \acrlong{ran}}
\newacronym{ca}{CA}{Carrier Aggregation}
\newacronym{cco}{CC}{Carrier Component}
\newacronym{nsa}{NSA}{Non Stand Alone}
\newacronym{embb}{eMBB}{Enhanced Mobility Broadband}
\newacronym{bsr}{BSR}{Buffer Status Report}
\newacronym{srb}{SRB}{Service Radio Bearer}
\newacronym{scm}{SCM}{Spatial Channel Model}
\newacronym{sctp}{SCTP}{Stream Control Transmission Protocol}
\newacronym{mptcp}{MPTCP}{Multi-path TCP}
\newacronym{ietf}{IETF}{Internet Engineering Task Force}
\newacronym{os}{OS}{Operating System}
\newacronym{tls}{TLS}{Transport Layer Security}
\newacronym{rfc}{RFC}{Request for Comments}
\newacronym{http}{HTTP}{HyperText Transfer Protocol}
\newacronym{nat}{NAT}{Network Address Translation}
\newacronym{api}{API}{Application Programming Interface}
\newacronym{rto}{RTO}{Retransmission Timeout}
\newacronym{psc}{PSC}{Public Safety Communication}
\newacronym{rpgm}{RPGM}{Reference Point Group Mobility}
\newacronym{ic}{IC}{Incident Command}
\newacronym{rsu}{RSU}{Road Side Unit}
\newacronym{uav}{UAV}{Unmanned Aerial Vehicle}
\newacronym{usv}{USV}{Unmanned Surface Vehicle}
\newacronym{uas}{UAS}{Unmanned Aerial System}
\newacronym{iab}{IAB}{Integrated Access and Backhaul}
\newacronym{qoe}{QoE}{Quality of Experience}
\newacronym{ssim}{SSIM}{Structural Similarity Index}
\newacronym{psnr}{PSNR}{Peak Signal to Noise Ratio}
\newacronym{bs}{BS}{Base Station}
\newacronym{mu}{MU}{Multiple User}
\newacronym{ag}{AG}{Air-to-Ground}

\usepackage{tikz}
\usepackage{pgfplots}
\pgfplotsset{compat=newest} 
\pgfplotsset{plot coordinates/math parser=false} 
\newlength\fheight
\newlength\fwidth
\usetikzlibrary{plotmarks,patterns,decorations.pathreplacing,backgrounds,calc,arrows,arrows.meta,spy,matrix}
\usepgfplotslibrary{patchplots,groupplots}
\usepackage{tikzscale}

\tikzstyle{startstop} = [rectangle, rounded corners, minimum width=2cm, minimum height=0.5cm,text centered, draw=black]
\tikzstyle{io} = [trapezium, trapezium left angle=70, trapezium right angle=110, minimum width=3cm, minimum height=1cm, text centered, draw=black]
\tikzstyle{process} = [rectangle, minimum width=2cm, minimum height=0.5cm, text centered, draw=black, align=center]
\tikzstyle{decision} = [ellipse, minimum width=2cm, minimum height=1cm, text centered, draw=black]
\tikzstyle{arrow} = [thick,<->,>=stealth]
\tikzstyle{line} = [thick,>=stealth]
\tikzstyle{darrow} = [thick,<->,>=stealth,dashed]
\tikzstyle{sarrow} = [thick,->,>=stealth]
\tikzstyle{larrow} = [line width=0.1mm,dashdotted,<->,>=stealth]

\definecolor{SchoolColor}{RGB}{0.71, 0, 0.106}
\definecolor{chaptergrey}{rgb}{0.61, 0, 0.09} 
\definecolor{midgrey}{rgb}{0.4, 0.4, 0.4}
\definecolor{chaptergreen}{rgb}{0.09, 0.612, 0}
\definecolor{chapterpurple}{rgb}{0.522, 0, 0.612}
\definecolor{chapterlightgreen}{rgb}{0, 0.612, 0.522}

\makeatletter
\def\grd@save@target#1{%
  \def\grd@target{#1}}
\def\grd@save@start#1{%
  \def\grd@start{#1}}
\tikzset{
  grid with coordinates/.style={
    to path={%
      \pgfextra{%
        \edef\grd@@target{(\tikztotarget)}%
        \tikz@scan@one@point\grd@save@target\grd@@target\relax
        \edef\grd@@start{(\tikztostart)}%
        \tikz@scan@one@point\grd@save@start\grd@@start\relax
        \draw[minor help lines] (\tikztostart) grid (\tikztotarget);
        \draw[major help lines] (\tikztostart) grid (\tikztotarget);
        \grd@start
        \pgfmathsetmacro{\grd@xa}{\the\pgf@x/1cm}
        \pgfmathsetmacro{\grd@ya}{\the\pgf@y/1cm}
        \grd@target
        \pgfmathsetmacro{\grd@xb}{\the\pgf@x/1cm}
        \pgfmathsetmacro{\grd@yb}{\the\pgf@y/1cm}
        \pgfmathsetmacro{\grd@xc}{\grd@xa + \pgfkeysvalueof{/tikz/grid with coordinates/major step x}}
        \pgfmathsetmacro{\grd@yc}{\grd@ya + \pgfkeysvalueof{/tikz/grid with coordinates/major step y}}
        \foreach \x in {\grd@xa,\grd@xc,...,\grd@xb}
        \node[anchor=north] at (\x,\grd@ya) {\pgfmathprintnumber{\x}};
        \foreach \y in {\grd@ya,\grd@yc,...,\grd@yb}
        \node[anchor=east] at (\grd@xa,\y) {\pgfmathprintnumber{\y}};
      }
    }
  },
  minor help lines/.style={
    help lines,
    gray,
    line cap =round,
    xstep=\pgfkeysvalueof{/tikz/grid with coordinates/minor step x},
    ystep=\pgfkeysvalueof{/tikz/grid with coordinates/minor step y}
  },
  major help lines/.style={
    help lines,
    line cap =round,
    line width=\pgfkeysvalueof{/tikz/grid with coordinates/major line width},
    xstep=\pgfkeysvalueof{/tikz/grid with coordinates/major step x},
    ystep=\pgfkeysvalueof{/tikz/grid with coordinates/major step y}
  },
  grid with coordinates/.cd,
  minor step x/.initial=.5,
  minor step y/.initial=.2,
  major step x/.initial=1,
  major step y/.initial=1,
  major line width/.initial=1pt,
}
\makeatother

\makeatletter

 \let\oldforeign@language\foreign@language
 \DeclareRobustCommand{\foreign@language}[1]{%
   \lowercase{\oldforeign@language{#1}}}

\usepackage[font=small]{caption}
\usepackage[font=footnotesize]{subcaption}

\usepackage{dblfloatfix}

\makeatother

\IEEEoverridecommandlockouts

\begin{document}


\title{Millimeter Wave Remote UAV Control and Communications for Public Safety Scenarios
}

\author{
%
\IEEEauthorblockN{William Xia$^{\dagger}$ \quad Michele Polese$^{\flat}$ \quad Marco Mezzavilla$^{\dagger}$ \quad  Giuseppe Loianno$^{\dagger}$ \quad Sundeep Rangan$^{\dagger}$ \quad Michele Zorzi$^{\flat}$} 
\IEEEauthorblockA{$^{\dagger}$NYU Tandon School of Engineering, Brooklyn, NY, USA}
\IEEEauthorblockA{$^{\flat}$Department of Information Engineering, University of Padova, 35131 Padua, Italy}

}

\maketitle

\begin{abstract}  
Communication and video capture from 
unmanned aerial vehicles (UAVs) offer significant
potential for assisting first responders
in remote public safety settings.
In such uses, millimeter wave (mmWave) wireless links 
can provide high throughput and low latency connectivity
between the UAV and a remote command center. 
However, maintaining reliable aerial 
communication in the mmWave bands 
is challenging due to the need to support high speed beam tracking 
and overcome blockage.  
This paper provides a simulation study aimed at assessing the feasibility of
public safety UAV connectivity through
a 5G link at 28~GHz.
Real flight motion traces are captured during maneuvers
similar to those expected in public safety settings.
The motions traces are then incorporated into
a detailed mmWave network simulator that models
the channel, blockage, beamforming and full 3GPP protocol stack.
We show that 5G mmWave communications can  deliver  throughput  up  to 1 Gbps  with  consistent  sub  ms latency when the base station is located near the mission area, enabling remote offloading of the UAV control and perception algorithms.

\end{abstract}

\begin{IEEEkeywords}
UAV, mmWave, remote control, performance evaluation
\end{IEEEkeywords}

\section{Introduction}
\label{sec:intro}

\glspl{uav} are being increasingly considered for use in public safety scenarios
\cite{merwaday2015uav,he2017drone,mezzavilla2017publicSafety6GHz}.  UAVs can provide valuable video and surveillance  
of emergency zones as well as wireless connectivity in regions where cellular infrastructure is either not
available (such as wildfires in remote areas) or where that infrastructure is no longer operational.
In such scenarios, high bandwidth, low-latency remote
communication to the UAVs may be of value.
For example, 
a UAV could provide wireless connectivity to a number of 
first responders in a disaster scenario and then backhaul that traffic to a command unit.
Since the UAV would be aggregating traffic from several sources and will have to exchange uncompressed data from high bit-rate sensors, such as advanced hyper-spectral cameras and lidars, the total bandwidth requirement can be much larger than what traditional frequency bands can actually offer. More specifically, in order to push off-board the heavy computational burden of the artificial intelligence algorithms used to process the data, Gbps of uncompressed data\footnote{Lidars themselves can generate millions of points per second with over 50 channels each \cite{lidars}.} will have to be uploaded in real time to the cloud or to the edge of the network~\cite{BARBAROSSA2018419}. 

In addition, remote control of the drone may demand very low latency connectivity, particularly
for high velocity and agile maneuvers of small scale vehicles navigating outside the line of sight~\cite{LoiannoRAL2017}, or scenarios with strong wind perturbations~\cite{LoiannoRAL2018_desert}.  

The massive bandwidth available in the millimeter wave (mmWave) bands 
offers the possibility of both very high throughput and low latency connectivity 
\cite{rangan2014millimeter,rappaport2013millimeter}.  Due to their enormous potential,
communication in the mmWave bands along with other transmissions above 6~GHz have emerged as a central 
component of the 3GPP Fifth Generation (5G) NR standard \cite{3GPP38.300}.
UAV communication is also already being extensively studied for 4G LTE 
\cite{muruganathan2018overview,3GPP36.777} and 5G NR may offer significantly greater capabilities.

Nevertheless, there are several key challenges in mmWave communications for UAVs:
due to high isotropic path loss, 
mmWave signals are transmitted in narrow, electrically steerable beams and reliable 
communication requires rapid
beam tracking and steering.  Such tracking can be a challenge in high velocity flight.
Besides, for low height scenarios, typically in the range of $10-20$ m,
line-of-sight links may be blocked by buildings and other obstacles in the environments, leading
to intermittent connectivity.
The broad purpose of this paper is twofold:  
(1) to assess throughput and latency requirements for the offloading at the edge of the control of drones, considering a public safety setting; 
and (2) evaluate the feasibility of mmWave connections for this task.

To perform this evaluation, 
we have collected detailed motion traces for actual drone flights in settings 
that we expect are similar to the maneuvers in public safety settings.  The motion traces include
position, velocity and orientation.  

\begin{figure}[t]
\centering
\includegraphics[width=1\columnwidth]{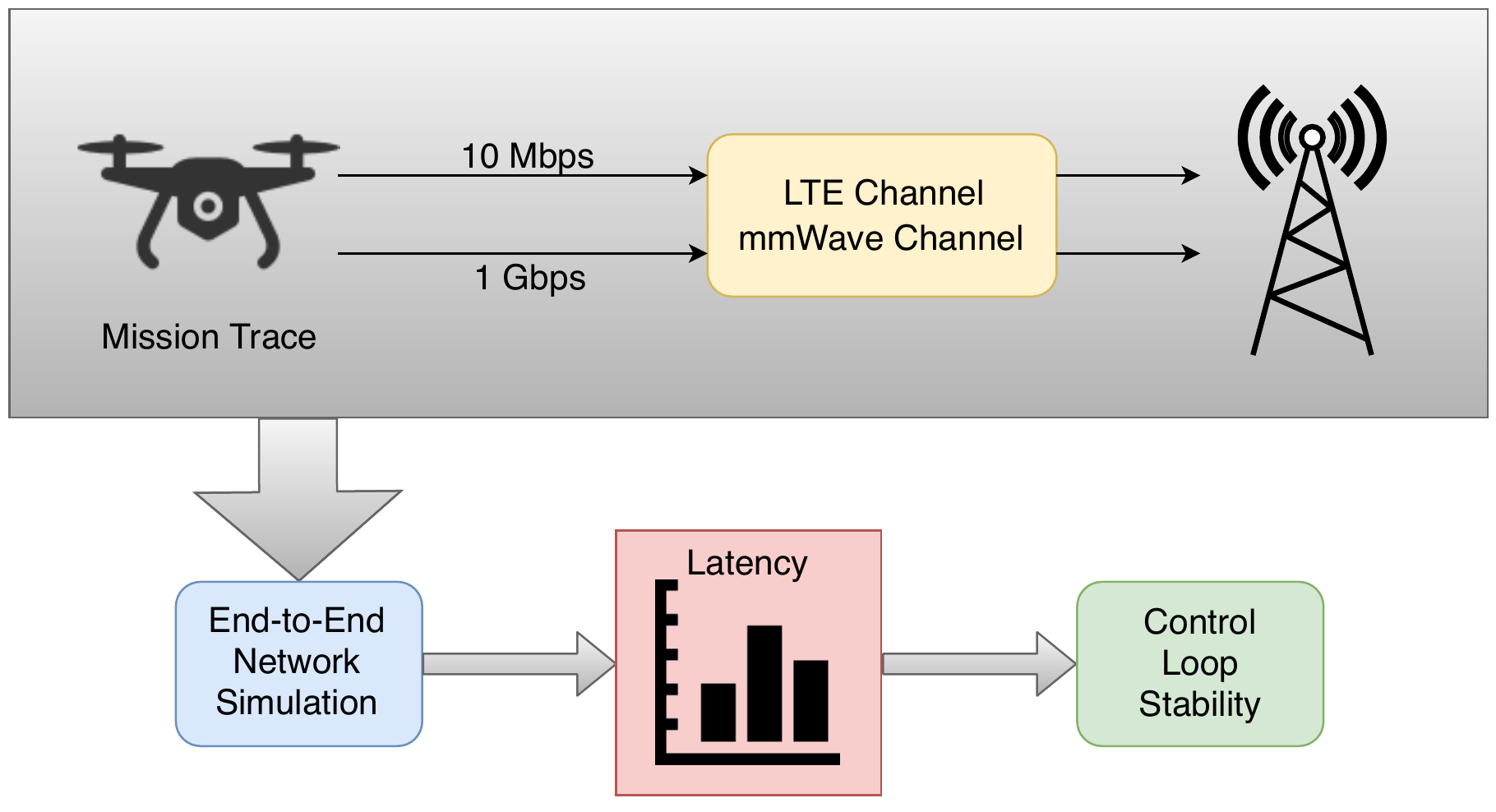}
\caption{Performance Evaluation Diagram.}
\label{fig:flowchart}
\end{figure}

We then simulate a hypothetical mmWave link in this flight using the detailed network simulator
described in \cite{mezzavilla2018end}.
Importantly, the simulation includes the effects of beamforming -- a key challenge in mmWave aerial communication.
The full upper layer protocol stack of 5G cellular networks is also modeled.  
The simulation then provides throughput trends and the effect of the latency can be
evaluated in the context of drone control offloading.  
The methodology of our approach is depicted in Fig.~\ref{fig:flowchart}.
To the best of our knowledge, this is the first comprehensive end-to-end simulation of 5G mmWave UAV communication and control.

Our initial results show that 5G mmWave communications can deliver throughput up to $1$ Gbps with consistent sub ms latency when the base station is located near the mission area. 

The remainder of the paper is organized as follows. We present a review of the state of the art in Sec.~\ref{sec:soa}, and discuss why mmWave communications can be beneficial for remote \gls{uav} control in Sec.~\ref{sec:uav}. Then, we describe the real flight traces and the associated missions in Sec.~\ref{sec:flights}, and present our results in Sec.~\ref{sec:results}. Finally, Sec.~\ref{sec:conclusions} concludes the paper and presents possible extensions.

\section{State of the Art}
\label{sec:soa}

In this section, we discuss the advances and work done in the area of wireless communications for \gls{uav} at mmWave frequencies. An overview and presentation of the benefits of using \gls{uav} for cellular networks is discussed in \cite{zeng2019UAVcellular}, while~\cite{xiao2016enabling} focuses on mmWave communications in aerial scenarios. The paper~\cite{mezzavilla2017publicSafety6GHz} explores the potentials of frequencies above 6 GHz for \gls{psc}, and provides a specific use case for mmWave communications with a \glspl{uav} in a wildfire scenario.

The numerical characterization of the performance of airborne mmWave networks depends on the modeling of the \gls{uav} air-to-ground and air-to-air channel.  \cite{khawaja2017mmWaveUAVChannel} investigates the characteristics of the mmWave air-to-ground channels for \gls{uav} in two frequency bands ($28$ GHz and $60$ GHz) through ray tracing and an experimental setup, showing that the received signal strength follows a two ray propagation model. Similarly,~\cite{8439182} analyzes the temporal and spatial characteristics of a 28 GHz air-to-ground channel through ray tracing, highlighting the impact of the scatterers on the overall channel behavior. 
\cite{cuverlier2018mmWaveAerial} presents a study of mmWave \gls{mu}-\gls{mimo} networks wherein atmospheric attenuation effects are modeled and single path channels are considered. Finally, channel tracking methods based on \gls{uav} movement state information and channel gain information are proposed by \cite{zhao2018mmWaveFlightControl}. Utilization and study of use cases for \glspl{uav} are discussed in \cite{khosravi2018UAVmmWaveOperation} and \cite{rupasinghe2017NonOrthonormal}. Both of these papers propose the use of \glspl{uav} as base stations and develop methods to optimize network coverage and spectral efficiency. 

With respect to the state of the art, this paper evaluates the end-to-end performance of \glspl{uav} at mmWave frequencies for real public safety flight mission scenarios, also considering the requirements for the offloading of the control algorithms of the \gls{uav} to the edge of the network. This scenario considers a uplink heavy connection to the \glspl{uav}, rather than the downlink heavy cases of the existing work related to the \gls{uav} deployment as base stations.

\section{Autonomy over 5G}
\label{sec:uav}
As mentioned in Sec.~\ref{sec:intro}, cellular or ad hoc mmWave deployments could be used to remotely control a \gls{uav}, and at the same time stream telemetry, video and user data from the \gls{uav} to a remote location. However, there exist bounds in the tolerable latency that should not be exceeded in order to maintain stability in the control of the system.

We discuss here the autonomy problem of a quadrotor vehicle in terms of control and perception, while the control loop is subject to delays in communication or computation. The main goal is to give the reader an overview about potential benefits of 5G and mmWave communications in this field.

The recent advances in computing, sensing, and perception algorithms provide the ability to run most of the autonomous components on board of the small scale aerial vehicles. Current controllers for quadrotor or, in general, Vertical Take-off and Landing (VTOL) commercial vehicles are composed of an inner loop and an outer loop. The inner loop is responsible for stabilizing the attitude dynamics of the platform, whereas the outer control loop is used to control the position of the robot. If the inner loop runs on the vehicle, then current communication technologies, such as Wi-Fi and LTE, already provide the ability to remotely run the position control on a ground station. Under these conditions, despite the system being slower, it is still possible to demonstrate the stability of the overall quadrotor system. To show this property, it is possible to model the system as a double integrator, with independent subsystems for each Cartesian component, and assuming that the inner attitude loop is faster than the outer one. The time scale separation between inner and outer loops is a common assumption in the literature~\cite{5569026}. In these conditions, considering modeling errors and algorithm delays due to perception processing algorithms (typically within $10-20$ ms), the average available phase margin is around $40$ deg.

\begin{figure*}[t]
\centering
    \begin{subfigure}[t]{.85\columnwidth} 
        \centering
        \includegraphics[clip,width=\columnwidth]{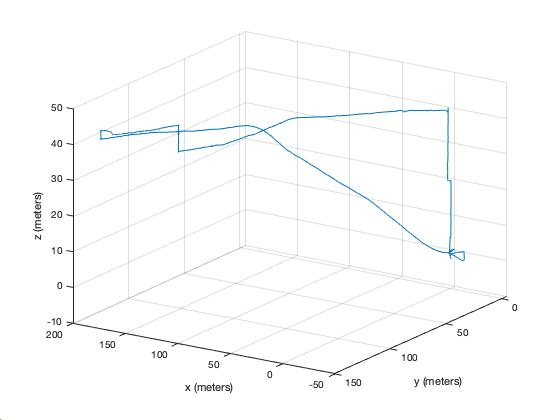}
        \caption{Crowd Overwatch (mission 1)}
        \label{fig:crowd}
    \end{subfigure}%
    \begin{subfigure}[t]{.85\columnwidth} 
        \centering
        \includegraphics[clip,width=\columnwidth]{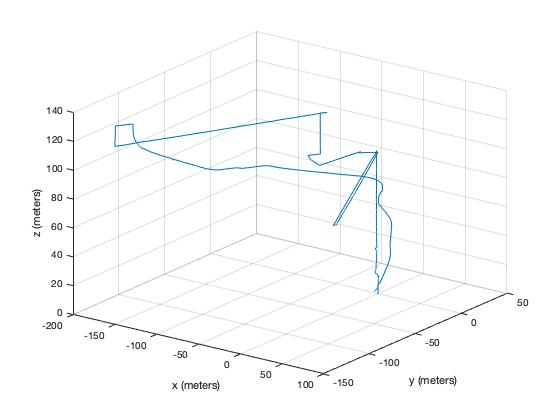}
        \caption{Missing Person (mission 2)}
        \label{fig:missing}
    \end{subfigure}
    \begin{subfigure}[t]{.85\columnwidth} 
        \centering
        \includegraphics[clip,width=\columnwidth]{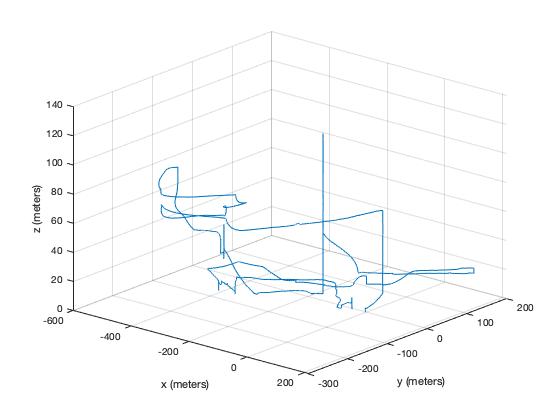}
        \caption{Prescribed Burn (mission 3)}
        \label{fig:burn}
    \end{subfigure}%
    \begin{subfigure}[t]{.85\columnwidth} 
        \centering
        \includegraphics[clip,width=\columnwidth]{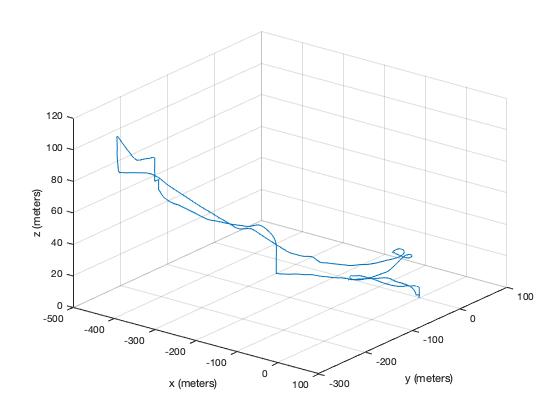}
        \caption{Sonar Boat Training (mission 4)}
        \label{fig:sonar}
    \end{subfigure}
    \setlength\belowcaptionskip{-.3cm}
    \caption{Plots of Drone Mission Flights.}
    \label{fig:mission_flights}
\end{figure*}

\begin{figure}[t!]
\centering
\includegraphics[width=0.78\columnwidth]{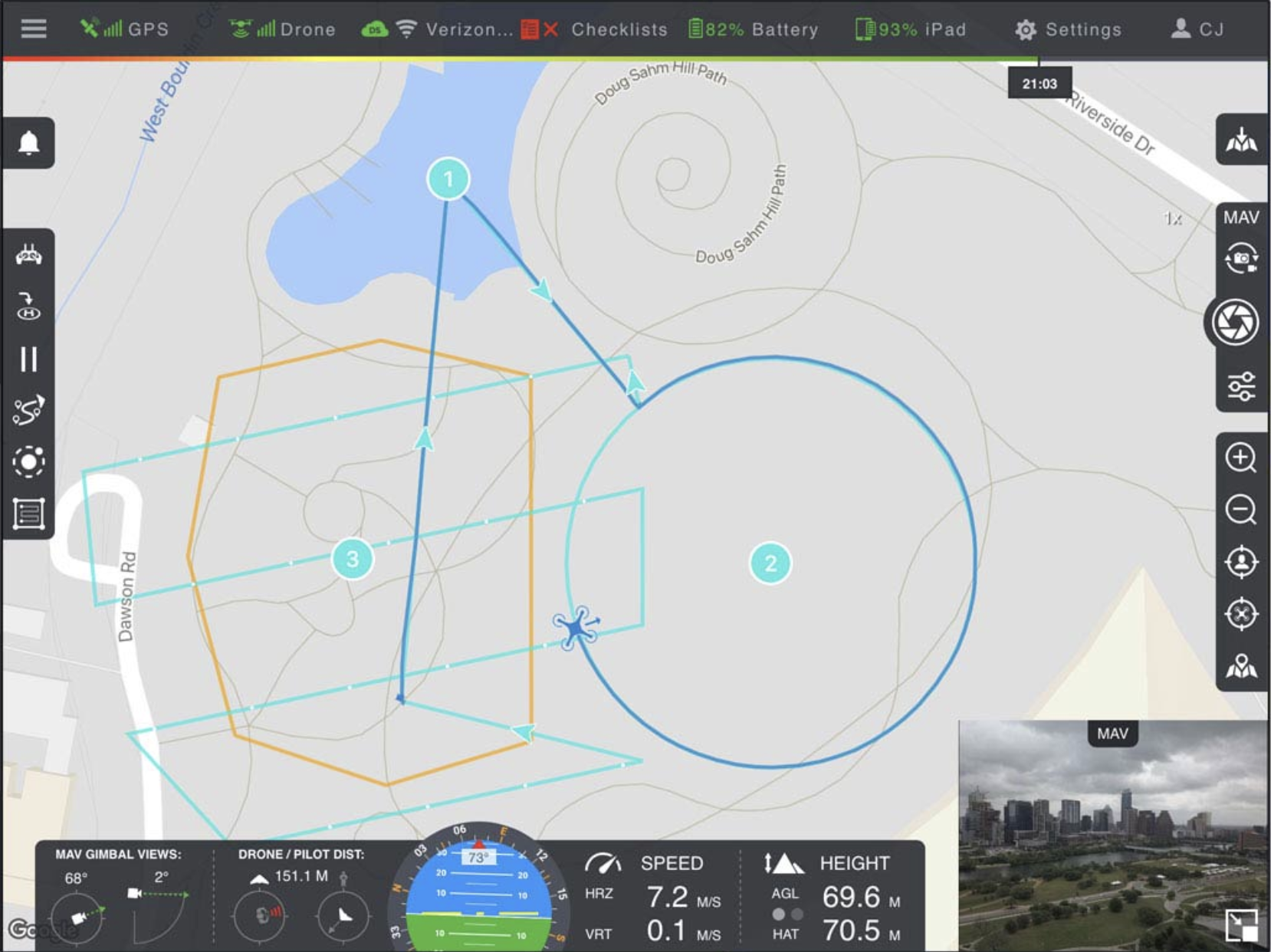}
\caption{Graphical interface of the software platform provided by DroneSense.}
\label{fig:dronesense}
\vspace{-5mm}
\end{figure}

Nonetheless, by using 5G and mmWave communications, it is possible to obtain a better performance, particularly in terms of control system reactiveness. The internal processing needed to solve the perception problem still takes around 10 to 20 ms~\cite{6906584,7487292}. By exploiting the sub-ms latency and the high bandwidth of mmWave links, it is possible to efficiently offload the computation to edge-based systems, e.g., in the \glspl{bs}, thus overcoming the limits in computational power of the on-board processing units and solving the perception problem in a shorter time interval. This provides more robustness and reactiveness to the autonomous vehicle. The usage of off-board processors has also the potential to reduce the scale and size of the platforms with the inherent additional benefit to increase users' safety.

\section{Drone Mission Flights}
\label{sec:flights}

The data we consider for our performance evaluation was provided by DroneSense, a company that partners up with our collaborators at \gls{afd} and develops a software platform for drones in public safety scenarios to expand situational awareness. The image in Fig.~\ref{fig:dronesense} represents the graphical interface provided by DroneSense's software platform, whereas the plots in Fig.~\ref{fig:mission_flights} show the flight paths each drone takes during its respective mission. 

The traces incorporate four large scale critical missions: 
\begin{itemize}
    \item \textbf{Crowd Overwatch:} The drones are used to monitor crowd migration, evacuation corridors and the gathering of situational awareness for the Incident Commander during a large outdoor music festival.
    \item \textbf{Missing Person:} The \gls{afd} \gls{red} team participates in the search efforts for missing persons in both the green space and throughout various lake and river networks. The drone traces provide a virtual timeline of the aircraft being used to locate a victim that was reported missing in a remote woodland area.
    \item \textbf{Prescribed Burn:} The \gls{afd} regularly participates in the activity of prescription or controlled burns in order to reduce vegetative fuel loads and prevent extreme wildfires. The drones are used to track personnel, monitor fire behavior, and provide the Burn Boss/Incident Commander critical information from an elevated vantage.
    \item \textbf{Sonar Boat Training:} The \gls{afd} recently received a \gls{usv} through the Department of Navy. This system is used to search for victims and other points of interest located below the water surface. The \gls{afd} \gls{red} team acknowledges the value of machine teaming to enhance the outcomes of using both the \gls{usv} and \gls{uas} in both training and real world deployments. The \gls{uas} traces demonstrate how drones can be used to track, monitor and visualize surface operations of the \gls{usv} aiding to the situational awareness required for emergency operations.
\end{itemize}

\section{End-to-end Performance Evaluation}
\label{sec:results}

We evaluated the performance of the \gls{uav} in terms of \gls{snr}, achievable throughput and latency using the open source ns-3 mmWave module described in~\cite{mezzavilla2018end}. This simulator features a full-stack implementation of 3GPP-like \glspl{bs} and \glspl{ue}, several channel models for mmWave propagation and fading, and procedures to handle beamforming and mobility events. Moreover, thanks to the integration with ns-3~\cite{henderson2008network}, it is it possible to integrate in the simulation the application, transport and network layers, and the modeling of the terminal mobility. This becomes particularly relevant for the simulation campaigns we present in this paper, where we consider a fixed \gls{bs}, close to the \gls{ic} station, and the remote terminal installed in a \gls{uav}, which moves according to the traces of the missions described in Sec.~\ref{sec:flights}.

\subsection{Flight Data Processing and Mobility Model} \label{data_processing}
During the flight missions, the flight controller in the \gls{uav} captures location data using a GPS-like format, i.e., with the latitude and the longitude of each visited point. We note that the geodesic curve on which any of these flight missions occur is nearly flat, given that the distance that the \gls{uav} spans is in the order of a few hundred meters, as shown in Fig.~\ref{fig:mission_flights}. Therefore, we convert the latitude and longitude measurements to 2-dimensional Cartesian coordinates by assuming a constant distance of $111$ km between latitude lines. We proceed similarly for longitude lines, considering the average distance they have at the latitude at which the missions took place. This approximation is made since any warping to the flight patterns is negligible for the motivations of this work.

We use the ns-3 \texttt{WaypointMobilityModel} to integrate the mobility traces for the four scenarios in the simulations. The model accepts a 4-dimensional vector as a \textit{waypoint}, with three spatial coordinates and a corresponding time stamp. Due to the complexity of the mobility model, the number of waypoints used for each simulation is only a fraction of the total coordinate points from the original extracted data, but is still representative of the mobility of the \gls{uav} during the flight. Notice that we do not account for the orientation of the vehicle; the modeling of this feature is left as a future work.

\begin{table}[t!]
    \caption{Main simulation parameters}
    \label{table:params}
    \begin{tabular}{ll}
    \toprule
    Parameter & Value \\ \midrule
    mmWave carrier frequency $f_c$ & 28 GHz \\
    mmWave bandwidth $B$ & 1 GHz \\
    LTE carrier frequency $f_c$ & 2.1 GHz \\
    LTE bandwidth $B$ & 20 MHz \\
    Beamforming update period & 5 ms \\
    Antenna combinations $A = N_{\rm BS} \times N_{\rm UAV}$ & $16 \times 4$, $64 \times 16$ \\
    Application source rate $R$ & 10, 1000 Mbps \\
    BS height & 25 m \\
    \bottomrule
    \end{tabular}
\end{table}

\subsection{Simulation Parameters}
For the mmWave links, the simulation uses the same channel model used for the evaluations in~\cite{mezzavilla2017publicSafety6GHz,Polese:2017:MFP:3152896.3152905}, i.e., a single-ray \gls{los} model with free space propagation, according to the 3GPP model in~\cite{38900}, shadowing, and the Doppler effect introduced by the mobility of the vehicle. We consider a bandwidth $B=1$ GHz with a carrier frequency $f_c=28$ GHz, as reported in Table~\ref{table:params}. The physical layer model in the simulator implements an adaptive modulation and coding scheme which yields a throughput of 3.2 Gbps for a \gls{bs} with the best channel conditions. We also compare the performance of the mmWave connection with a baseline architecture with LTE connectivity at frequency below 6 GHz (i.e., $f_c = 2.1$ GHz). While, in general, the LTE connection enjoys a reduced pathloss with respect to mmWave, it can only exploit a much smaller bandwidth, which generally yields a much lower throughput. Moreover, the frame design of LTE does not support the sub-ms latency of the mmWave connection~\cite{ford2017latency}.

When considering mmWave communications in the context of \gls{uav} mobility, a critical operation is the beam tracking, which makes it possible to maintain a fine alignment of the transmitter and receiver beams and obtain a high beamforming gain~\cite{xiao2016enabling}. While beam management operations are fundamental also in cellular networks~\cite{giordani2019tutorial}, efficient and precise tracking is harder to obtain with flying vehicles, which present more erratic mobility patterns. Therefore, when simulating operations with \glspl{uav} at mmWave frequencies, it is important to model a realistic beam tracking procedure. In this paper, following the approach presented in~\cite{mezzavilla2017publicSafety6GHz}, we model the beam tracking by updating the beam pair to the optimal one with a fixed periodicity, and by keeping it fixed in the time interval between two consecutive updates. The periodicity we consider (5 ms) is compatible with that of the transmission of tracking reference signals (e.g., \glspl{csirs}) in 3GPP NR~\cite{giordani2019tutorial}.

For the application, we consider a source that periodically generates a payload of 1500 bytes, for a total source rate of 10 or 1000 Mbps, in order to model different use cases (e.g., sensor information, video at different rates). We consider UDP as transport protocol, and the transmissions are in uplink, i.e., from the \gls{uav} to the \gls{ic} station.  

\renewcommand{\thefigure}{6}
\begin{figure*}[b]
\centering
    \begin{subfigure}[t]{0.5\textwidth} 
        \centering
        \setlength\fwidth{0.75\columnwidth}
        \setlength\fheight{0.35\columnwidth}
%
%
\definecolor{mycolor1}{rgb}{0.00000,0.44700,0.74100}%
\definecolor{mycolor2}{rgb}{0.85000,0.32500,0.09800}%
\definecolor{mycolor3}{rgb}{0.92900,0.69400,0.12500}%

\definecolor{lavender}{rgb}{0.9020,0.9020,0.9804}%
\definecolor{lightskyblue}{rgb}{0.6784,0.8471,0.9020}%
\definecolor{deepskyblue}{rgb}{0,0.7490,1}%
\definecolor{steelblue}{rgb}{0.2745,0.5098,0.7059}%
\definecolor{blue}{rgb}{0,0,1}%
\definecolor{royalblue}{rgb}{0.2549,0.4118,0.8824}%

\definecolor{gainsboro}{rgb}{0.8627,0.8627,0.8627}%
\definecolor{darkslategrey}{rgb}{0.1843,0.3098,0.3098}%
\definecolor{gray}{rgb}{0.5,0.5,0.5}%

\definecolor{lightcoral}{rgb}{0.9412,0.5020,0.5020}%
\definecolor{indianred}{rgb}{0.8039,0.3608,0.3608}%
\definecolor{lightsalmon}{rgb}{1.0000,0.6275,0.4784}%
\definecolor{darksalmon}{rgb}{0.9137,0.5882,0.4784}%
\begin{tikzpicture}
\pgfplotsset{every tick label/.append style={font=\scriptsize}}

\begin{axis}[%
width=0.951\fwidth,
height=\fheight,
at={(0\fwidth,0\fheight)},
scale only axis,
bar shift auto,
xmin=0.5,
xmax=4.5,
xtick={1, 2, 3, 4},
xlabel style={font=\scriptsize\color{white!15!black}},
xlabel={Mission},
ymin=0,
ymax=1050,
ylabel style={font=\scriptsize\color{white!15!black}},
ylabel={Throughput [Mbps]},
axis background/.style={fill=white},
xmajorgrids,
ymajorgrids,
legend columns=3,
legend style={at={(1.25,1.05)}, anchor=south, font=\scriptsize, legend cell align=left, align=left, draw=white!15!black}
]
\addplot[ybar, bar width=0.178, fill=mycolor1, draw=black, area legend] table[row sep=crcr] {%
1	1015.47861250655\\
2	1009.83584713863\\
3	783.203328104242\\
4	950.640855538462\\
};
\addplot[forget plot, color=white!15!black] table[row sep=crcr] {%
0.5	0\\
4.5	0\\
};
\addlegendentry{mmWave link, $A=16\times4$}

\addplot[ybar, bar width=0.178, fill=mycolor2, draw=black, area legend] table[row sep=crcr] {%
1	1019.2177196846\\
2	1016.08412881838\\
3	1018.27434390667\\
4	1016.68143058974\\
};
\addplot[forget plot, color=white!15!black] table[row sep=crcr] {%
0.5	0\\
4.5	0\\
};
\addlegendentry{mmWave link, $A=64\times16$}

\addplot[ybar, bar width=0.178, fill=mycolor3, draw=black, area legend] table[row sep=crcr] {%
1	75.2125654426231\\
2	75.1661036123826\\
3	75.2262563762041\\
4	75.0230008461542\\
};
\addplot[forget plot, color=white!15!black] table[row sep=crcr] {%
0.5	0\\
4.5	0\\
};
\addlegendentry{LTE link}

\end{axis}
\end{tikzpicture}%
        \caption{Throughput}
        \label{fig:m_th}
    \end{subfigure}%
    \begin{subfigure}[t]{0.5\textwidth} 
        \centering
        \setlength\fwidth{0.75\columnwidth}
        \setlength\fheight{0.35\columnwidth}
%
%
\definecolor{mycolor1}{rgb}{0.00000,0.44700,0.74100}%
\definecolor{mycolor2}{rgb}{0.85000,0.32500,0.09800}%
\definecolor{mycolor3}{rgb}{0.92900,0.69400,0.12500}%

\definecolor{lavender}{rgb}{0.9020,0.9020,0.9804}%
\definecolor{lightskyblue}{rgb}{0.6784,0.8471,0.9020}%
\definecolor{deepskyblue}{rgb}{0,0.7490,1}%
\definecolor{steelblue}{rgb}{0.2745,0.5098,0.7059}%
\definecolor{blue}{rgb}{0,0,1}%
\definecolor{royalblue}{rgb}{0.2549,0.4118,0.8824}%

\definecolor{gainsboro}{rgb}{0.8627,0.8627,0.8627}%
\definecolor{darkslategrey}{rgb}{0.1843,0.3098,0.3098}%
\definecolor{gray}{rgb}{0.5,0.5,0.5}%

\definecolor{lightcoral}{rgb}{0.9412,0.5020,0.5020}%
\definecolor{indianred}{rgb}{0.8039,0.3608,0.3608}%
\definecolor{lightsalmon}{rgb}{1.0000,0.6275,0.4784}%
\definecolor{darksalmon}{rgb}{0.9137,0.5882,0.4784}%
\begin{tikzpicture}
\pgfplotsset{every tick label/.append style={font=\scriptsize}}

\begin{semilogyaxis}[%
width=0.951\fwidth,
height=\fheight,
at={(0\fwidth,0\fheight)},
scale only axis,
bar shift auto,
xmin=0.5,
xmax=4.5,
xtick={1, 2, 3, 4},
xlabel style={font=\scriptsize\color{white!15!black}},
xlabel={Mission},
ymin=0,
ymax=120,
ylabel style={font=\scriptsize\color{white!15!black}},
ylabel={RAN latency [ms]},
axis background/.style={fill=white},
xmajorgrids,
ymajorgrids,
legend style={legend cell align=left, align=left, draw=white!15!black, font=\scriptsize}
]
\addplot[ybar, bar width=0.178, fill=mycolor1, draw=black, area legend] table[row sep=crcr] {%
1	0.301921561783271\\
2	11.9938205038222\\
3	1.85000515483315\\
4	1.59434660240817\\
};
\addplot[forget plot, color=white!15!black] table[row sep=crcr] {%
0.5	0\\
4.5	0\\
};

\addplot[ybar, bar width=0.178, fill=mycolor2, draw=black, area legend] table[row sep=crcr] {%
1	0.294061765884937\\
2	0.44957208949976\\
3	0.318338872045662\\
4	0.35123300511995\\
};
\addplot[forget plot, color=white!15!black] table[row sep=crcr] {%
0.5	0\\
4.5	0\\
};

\addplot[ybar, bar width=0.178, fill=mycolor3, draw=black, area legend] table[row sep=crcr] {%
1	116.117754956253\\
2	116.117754956253\\
3	116.117754956253\\
4	116.117754956253\\
};
\addplot[forget plot, color=white!15!black] table[row sep=crcr] {%
0.5	0\\
4.5	0\\
};

\end{semilogyaxis}
\end{tikzpicture}%
        \caption{Latency (the y axis is in logarithmic scale)}
        \label{fig:m_lat}
    \end{subfigure}
    \caption{Performance for different missions, with different antenna combinations and a source rate of 1000 Mbps.}
    \label{fig:missions_1gbps}
\end{figure*}

\renewcommand{\thefigure}{7}
\begin{figure*}[b]
\centering
    \begin{subfigure}[t]{0.5\textwidth} 
        \centering
        \setlength\fwidth{0.75\columnwidth}
        \setlength\fheight{0.35\columnwidth}
%
%
\definecolor{mycolor1}{rgb}{0.00000,0.44700,0.74100}%
\definecolor{mycolor2}{rgb}{0.85000,0.32500,0.09800}%
\definecolor{mycolor3}{rgb}{0.92900,0.69400,0.12500}%
\definecolor{lavender}{rgb}{0.9020,0.9020,0.9804}%
\definecolor{lightskyblue}{rgb}{0.6784,0.8471,0.9020}%
\definecolor{deepskyblue}{rgb}{0,0.7490,1}%
\definecolor{steelblue}{rgb}{0.2745,0.5098,0.7059}%
\definecolor{blue}{rgb}{0,0,1}%
\definecolor{royalblue}{rgb}{0.2549,0.4118,0.8824}%

\definecolor{gainsboro}{rgb}{0.8627,0.8627,0.8627}%
\definecolor{darkslategrey}{rgb}{0.1843,0.3098,0.3098}%
\definecolor{gray}{rgb}{0.5,0.5,0.5}%

\definecolor{lightcoral}{rgb}{0.9412,0.5020,0.5020}%
\definecolor{indianred}{rgb}{0.8039,0.3608,0.3608}%
\definecolor{lightsalmon}{rgb}{1.0000,0.6275,0.4784}%
\definecolor{darksalmon}{rgb}{0.9137,0.5882,0.4784}%
\begin{tikzpicture}
\pgfplotsset{every tick label/.append style={font=\scriptsize}}

\begin{axis}[%
width=0.951\fwidth,
height=\fheight,
at={(0\fwidth,0\fheight)},
scale only axis,
bar shift auto,
xmin=0.5,
xmax=4.5,
xtick={1, 2, 3, 4},
xlabel style={font=\scriptsize\color{white!15!black}},
xlabel={Mission},
ymin=0,
ymax=12,
ylabel style={font=\scriptsize\color{white!15!black}},
ylabel={Throughput [Mbps]},
axis background/.style={fill=white},
xmajorgrids,
ymajorgrids,
legend columns=3,
legend style={at={(1.25,1.05)}, anchor=south, font=\scriptsize, legend cell align=left, align=left, draw=white!15!black}
]
\addplot[ybar, bar width=0.178, fill=mycolor1, draw=black, area legend] table[row sep=crcr] {%
1	10.1923356959763\\
2	10.1861018896367\\
3	10.1941239796033\\
4	10.1669290512821\\
};
\addplot[forget plot, color=white!15!black] table[row sep=crcr] {%
0.5	0\\
4.5	0\\
};
\addlegendentry{mmWave link, $A=16\times4$}

\addplot[ybar, bar width=0.178, fill=mycolor2, draw=black, area legend] table[row sep=crcr] {%
1	10.1923356959763\\
2	10.1861018896367\\
3	10.1941725235127\\
4	10.1669290512821\\
};
\addplot[forget plot, color=white!15!black] table[row sep=crcr] {%
0.5	0\\
4.5	0\\
};
\addlegendentry{mmWave link, $A=64\times16$}

\addplot[ybar, bar width=0.178, fill=mycolor3, draw=black, area legend] table[row sep=crcr] {%
1	10.1922990938898\\
2	10.1860357792733\\
3	10.1941446934844\\
4	10.1667716153846\\
};
\addplot[forget plot, color=white!15!black] table[row sep=crcr] {%
0.5	0\\
4.5	0\\
};
\addlegendentry{LTE link}

\end{axis}
\end{tikzpicture}%
        \caption{Throughput}
        \label{fig:m_th_10}
    \end{subfigure}%
    \begin{subfigure}[t]{0.5\textwidth} 
        \centering
        \setlength\fwidth{0.75\columnwidth}
        \setlength\fheight{0.35\columnwidth}
%
%
\definecolor{mycolor1}{rgb}{0.00000,0.44700,0.74100}%
\definecolor{mycolor2}{rgb}{0.85000,0.32500,0.09800}%
\definecolor{mycolor3}{rgb}{0.92900,0.69400,0.12500}%
\begin{tikzpicture}
\pgfplotsset{every tick label/.append style={font=\scriptsize}}

\begin{axis}[%
width=0.951\fwidth,
height=\fheight,
at={(0\fwidth,0\fheight)},
scale only axis,
bar shift auto,
xmin=0.5,
xmax=4.5,
xtick={1, 2, 3, 4},
xlabel style={font=\scriptsize\color{white!15!black}},
xlabel={Mission},
ymin=0,
ymax=6,
ylabel style={font=\scriptsize\color{white!15!black}},
ylabel={Latency [ms]},
axis background/.style={fill=white},
xmajorgrids,
ymajorgrids,
legend style={legend cell align=left, align=left, draw=white!15!black, font=\scriptsize}
]
\addplot[ybar, bar width=0.178, fill=mycolor1, draw=black, area legend] table[row sep=crcr] {%
1	0.481159699240137\\
2	0.456214833389938\\
3	0.51679707139331\\
4	0.486888118464882\\
};
\addplot[forget plot, color=white!15!black] table[row sep=crcr] {%
0.5	0\\
4.5	0\\
};

\addplot[ybar, bar width=0.178, fill=mycolor2, draw=black, area legend] table[row sep=crcr] {%
1	0.40957600000352\\
2	0.423384180452397\\
3	0.435288459437524\\
4	0.443578373753099\\
};
\addplot[forget plot, color=white!15!black] table[row sep=crcr] {%
0.5	0\\
4.5	0\\
};

\addplot[ybar, bar width=0.178, fill=mycolor3, draw=black, area legend] table[row sep=crcr] {%
1	5.32856999999115\\
2	5.32856999999278\\
3	5.32856972787367\\
4	5.32857077052072\\
};
\addplot[forget plot, color=white!15!black] table[row sep=crcr] {%
0.5	0\\
4.5	0\\
};

\end{axis}
\end{tikzpicture}%
        \caption{Latency}
        \label{fig:m_lat_10}
    \end{subfigure}
    \caption{Performance for different missions, with different antenna combinations and a source rate of 10 Mbps.}
    \label{fig:missions_10mbps}
\end{figure*}

\renewcommand{\thefigure}{4}
\begin{figure}[t]
    \setlength\fwidth{0.8\columnwidth}
    \setlength\fheight{0.4\columnwidth}
    \input{figures/snr-3.tex}
    \caption{SNR over time, for mission 3 (prescribed burn).}
    \label{fig:snr-time}
\end{figure}

\renewcommand{\thefigure}{5}
\begin{figure}[t]
    \setlength\fwidth{0.8\columnwidth}
    \setlength\fheight{0.4\columnwidth}
    \input{figures/lat-3.tex}
    \caption{Latency over time, for mission 3 (prescribed burn). The latency is averaged over fixed intervals of 5 seconds.}
    \label{fig:lat-time}
\end{figure}

\subsection{Results}

In this paragraph, we will report the main results of our preliminary performance evaluation, where we focus on three main metrics: (i) the \gls{snr} of the link; (ii) the throughput and (iii) the one-way latency. Both throughput and latency are measured at the \gls{pdcp} layer of the cellular protocol stack, given that we assume that the sink of the communications is at the \gls{ic} station and thus there is no additional delay introduced by forwarding over the public Internet.

Fig.~\ref{fig:snr-time} reports, as an example, the evolution of the \gls{snr} over time in mission 3, i.e., the prescribed burn, for the mmWave link and different combinations of the transmitter and receiver side antennas. Notice that, given that the \gls{bs} is located in the area of the mission of the \gls{uav}, the \gls{snr} is high for most of the mission duration also with the antennas with fewer elements, which in general introduce a lower beamforming gain. Nonetheless, a higher number of antenna elements can help to sustain a better connection in particularly critical conditions, e.g., when the \gls{uav} and the \gls{bs} at the \gls{ic} station are further away. 
This is shown in Fig.~\ref{fig:snr-time}, where it can be seen that for the three time intervals roughly between $t=580$ and 700 seconds, around $t=1000$ seconds and between $t=1500$ and 1600 seconds the \gls{snr} achievable with $A = 64 \times 16$ is approximately 10 dB higher than that with $A = 16 \times 4$. This has a direct impact on the latency of the connection, as shown in Fig.~\ref{fig:lat-time}: a lower SNR translates into a higher number of errors at the physical layer, and, consequently, retransmissions at the higher layers.

In order to better understand the limitations of the LTE connection and the potentials of mmWave communications for remote \gls{uav} control and data exchange, we plot in Figs.~\ref{fig:missions_1gbps} and~\ref{fig:missions_10mbps}  the average throughput and latency over the duration of each mission. In particular, we report results for the highest source rate we consider (i.e., 1000 Mbps) in Fig.~\ref{fig:missions_1gbps}, and for 10 Mbps in Fig.~\ref{fig:missions_10mbps}, and analyze the performance of the 2.1 GHz LTE link with that of the two antenna combinations for mmWave frequencies at 28 GHz. By comparing the two figures, it is possible to understand the trade-off between LTE and mmWaves. As expected, LTE can yield a low latency (i.e., below 10 ms, in line with the frame structure that is considered in this simulator~\cite{latencyreduction2017}) only with low souce rates, and cannot support rates higher than tens of Mpbs in uplink, as shown by Fig.~\ref{fig:m_th}. Therefore, the improved remote control of a \gls{uav} through offloading could be enabled only if the link is not used for the transmission of other kinds of data, and for a limited number of \glspl{uav}. Furthermore, this prevents for example the sharing of high-quality aerial video or other sensor data (e.g., LIDAR data) that cannot consequently be used to refine the performance of the control algorithms. The mmWave connection, instead, performs better in terms of both throughput and latency. Thanks to the high availability of bandwidth, the mmWave links manage to deliver the full 1000 Mbps throughput in almost all configurations. Moreover, the low-latency frame design that is coupled with mmWave communications in 5G cellular networks~\cite{giordani2019tutorial,dutta2017frame} helps achieve a small latency, with a sub-ms one-way latency that can enable a prompt control of the remote \gls{uav}, also if the connection is shared with data-intensive streams from, for example, high-resolution cameras.

When comparing different missions, it can be seen that for missions 1 and 2 both the antenna configurations can support the source rate of 1000 Mbps, while the full throughput is reached for missions 3 and 4 only with the largest antenna arrays tested. This configuration also provides the best performance in terms of latency.

\renewcommand{\thefigure}{8}
\begin{figure}[t]
    \setlength\fwidth{0.8\columnwidth}
    \setlength\fheight{0.5\columnwidth}
%
%
%
\definecolor{mycolor1}{rgb}{0.00000,0.44700,0.74100}%
\definecolor{mycolor2}{rgb}{0.85000,0.32500,0.09800}%
\definecolor{mycolor3}{rgb}{0.92900,0.69400,0.12500}%

\definecolor{lavender}{rgb}{0.9020,0.9020,0.9804}%
\definecolor{lightskyblue}{rgb}{0.6784,0.8471,0.9020}%
\definecolor{deepskyblue}{rgb}{0,0.7490,1}%
\definecolor{steelblue}{rgb}{0.2745,0.5098,0.7059}%
\definecolor{blue}{rgb}{0,0,1}%
\definecolor{royalblue}{rgb}{0.2549,0.4118,0.8824}%

\definecolor{gainsboro}{rgb}{0.8627,0.8627,0.8627}%
\definecolor{darkslategrey}{rgb}{0.1843,0.3098,0.3098}%
\definecolor{gray}{rgb}{0.5,0.5,0.5}%

\definecolor{lightcoral}{rgb}{0.9412,0.5020,0.5020}%
\definecolor{indianred}{rgb}{0.8039,0.3608,0.3608}%
\definecolor{lightsalmon}{rgb}{1.0000,0.6275,0.4784}%
\definecolor{darksalmon}{rgb}{0.9137,0.5882,0.4784}%
\begin{tikzpicture}
\pgfplotsset{every tick label/.append style={font=\scriptsize}}

\begin{axis}[%
width=0.951\fwidth,
height=\fheight,
at={(0\fwidth,0\fheight)},
scale only axis,
bar shift auto,
xmin=0.5,
xmax=2.5,
xtick=data,
xticklabels={Distant BS, On-premise BS},
xlabel style={font=\scriptsize\color{white!15!black}},
xlabel={BS location},
ymin=0,
ymax=24,
ylabel style={font=\scriptsize\color{white!15!black}},
ylabel={Latency [ms]},
axis background/.style={fill=white},
xmajorgrids,
ymajorgrids,
legend style={at={(0.98,0.98)}, anchor=north east, font=\scriptsize, legend cell align=left, align=left, draw=white!15!black}
]
\addplot[ybar, bar width=0.178, fill=mycolor1, draw=black, area legend] table[row sep=crcr] {%
1	23.0975\\
2	1.3035\\
};
\addlegendentry{$A=16\times4$}

\addplot[ybar, bar width=0.178, fill=mycolor2, draw=black, area legend] table[row sep=crcr] {%
1	7.7774\\
2	0.5524\\
};
\addlegendentry{$A=64\times16$}

\end{axis}
\end{tikzpicture}%
    \caption{Comparison of the average latency with different locations of the \gls{bs} covering the scenario: the distant \gls{bs} is 2 km away from the mission area, while the on-premise \gls{bs} is in the mission area.}
    \label{fig:lat-distance}
\end{figure}
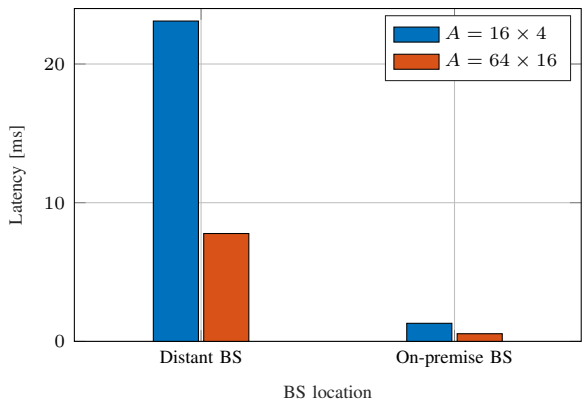

The same trend can be observed in Fig.~\ref{fig:lat-distance}, where we compare the average latency with a \gls{bs} on-premise, i.e., in the area of the mission (as for the previous results), and one placed 2 km apart. The rate we consider in this case is 1000 Mbps, which is not sustainable by the LTE connection. It can be seen that, despite the challenging communication condition introduced by the high propagation loss, the antenna configuration with $A = 64 \times 16$ is still capable of ensuring low-latency data exchange, and thus enable prompt and reactive \gls{uav} control also at larger distances. Moreover, as future work, we will explore the possibility of using an even larger number of antennas at the base station, to provide a higher beamforming gain and reduce even further the latency in the distant \gls{bs} scenario.

\section{Conclusions}
\label{sec:conclusions}
MmWave communications are a valuable resource in various realistic public safety scenarios, as described in Sec. \ref{sec:flights}. 
We offered an analysis of \gls{e2e} simulations and showed the potential of 5G networks to deliver throughput up to $1$ Gbps with consistent sub ms latency when the base station is located near the mission area. The benefits derived from the advent of 5G technology will certainly not be limited better controllability and reactivity of the aerial vehicles. In general, 5G signal traces have the potential to be used in a cooperative localization framework to identify the spatial location of different vehicles in GPS denied environments out of the line of sight of an operator, thereby increasing autonomy, safety, and usability of these platforms. Finally, the ability to outsource most of the computation to the edge cloud will also allow to scale down these vehicles, increasing user and environment safety.

That said, there are several extensions to this work that require attention in order to present this emerging technology as a complete and viable solution in the context of \glspl{psc}. 
As mentioned in Sec. \ref{sec:results}, the orientation of the \gls{uav} is not modeled. This factor is expected to significantly affect the beam tracking and thus a more complex simulation framework is needed to evaluate and understand these effects. 
Furthermore, we note that the environment for many public safety scenarios will likely have buildings within the mission area as well as wind perturbations. This will require simulations that model these blockage effects and perturbations in the flight path of the \gls{uav}. The throughput, latency and subsequent implications for the platform's stability can then be explored in more detail.

\section{Acknowledgments} We would like to thank Eric Schank (DroneSense) and Coitt Kessler (\gls{afd}) for their tremendous support. 

This work was partially supported by the U.S. Department of Commerce/NIST (Award No. 70NANB17H166), by the CloudVeneto initiative, and by the ARL grant DCIST CRA W911NF-17-2-0181.


\bibliographystyle{IEEEtran}
\bibliography{bibl}

\end{document}